\documentclass[reprint, superscriptaddress, amsmath, amssymb, aip, apl]{revtex4-2}

\usepackage{graphicx} 
\usepackage{xcolor}
\usepackage[colorlinks=true, allcolors=blue]{hyperref}

\newcommand{\wnmbr}[0]{\mathrm{cm}^{-1}}
\newcommand{\um}[0]{\mu\mathrm{m}}

\begin{document}

\preprint{}

\title{Interplay between growth conditions and formation of localized light emitters in epitaxial AlN thin films} 

\author{Meysam Saeedi}
\affiliation{Paul-Drude-Institut für Festkörperelektronik, Leibniz-Institut im Forschungsverbund Berlin e.V., Hausvogteiplatz 5-7, 10117 Berlin, Germany}

\author{Duc V. Dinh}
\affiliation{Paul-Drude-Institut für Festkörperelektronik, Leibniz-Institut im Forschungsverbund Berlin e.V., Hausvogteiplatz 5-7, 10117 Berlin, Germany}

\author{Andrey N. Anisimov}
\affiliation{Helmholtz-Zentrum Dresden-Rossendorf, Institute of Ion Beam Physics and Materials Research, Bautzner Landstrasse 400, 01328 Dresden, Germany}

\author{Mingyun Yuan}
\affiliation{Paul-Drude-Institut für Festkörperelektronik, Leibniz-Institut im Forschungsverbund Berlin e.V., Hausvogteiplatz 5-7, 10117 Berlin, Germany}

\author{Georgy V. Astakhov}
\affiliation{Helmholtz-Zentrum Dresden-Rossendorf, Institute of Ion Beam Physics and Materials Research, Bautzner Landstrasse 400, 01328 Dresden, Germany}

\author{Alberto Hernández-Mínguez}
\email{hernandez-minguez@pdi-berlin.de}
\affiliation{Paul-Drude-Institut für Festkörperelektronik, Leibniz-Institut im Forschungsverbund Berlin e.V., Hausvogteiplatz 5-7, 10117 Berlin, Germany}

\date{\today}

\begin{abstract}

AlN is a promising material for integrated quantum photonics due to its broad transparency window, its compatibility with CMOS technology, and the possibility to host localized light emitters acting as single photon sources. However, little is understood yet about the microscopic nature and formation mechanisms of these light emitters.
In this contribution, we demonstrate that their formation efficiency depends significantly on the growth conditions of the AlN thin film. We reveal that localized light emitters with distinct zero-phonon lines in the 600--800~nm range can be efficiently formed in epitaxial AlN thin films grown under N-rich conditions by exposing the AlN to a post-growth treatment consisting in a proton irradiation followed by an annealing process. In contrast, the density of light emitters formed in AlN grown under Al-rich conditions remains low, even after the thin films have been exposed to the post-growth treatment.
These results open a pathway toward a better understanding of the microscopic nature of these light emission centers, as well as their engineered integration as quantum light sources in AlN-based optomechanical platforms.

\end{abstract}

\maketitle


Aluminum nitride (AlN) is a promising platform for applications in ultraviolet optoelectronics due to its ultra wide band gap ($\approx$ 6.2~eV),\cite{yim_epitaxially_1973} as well as for integrated optomechanical applications due to its piezoelectricity and excellent electromechanical properties.\cite{xiong2012integrated}

Recently, it has been demonstrated that AlN can also host spatially localized light emitters with emission energies well within its broad transparency window acting as single photon sources.\cite{xue2020single, bishop_room-temperature_2020, lu2020bright, xue_experimental_2021, cannon_polarization_2023, wang_quantum_2023} This capability, combined with a substantial nonlinear optical susceptibility,\cite{fujii_nonlinear_1977} moderate electro-optic response,\cite{liu2023aluminum} and compatibility with CMOS technology,\cite{li2021aluminium, gundougdu2025algan} has also positioned AlN as a promising material for integrated quantum photonics in the visible to near-infrared wavelength range.

Although theoretical studies have tentatively attributed these spatially localized light emitters to intrinsic defect complexes,\cite{xue2020single, zhu_formation_2024} their microscopic nature and formation mechanisms still remain unclear. On the one hand, light emitters have been observed in pristine AlN thin films,\cite{xue2020single, pezzagna_photoinduced_2026} indicating that they can form spontaneously during AlN growth. On the other hand, several works have demonstrated that they can also be deliberately formed via post-growth processing techniques.\cite{lu2020bright, wang_quantum_2023, nieto_hernandez_fabrication_2024, senichev_quantum_2024}

In this contribution, we demonstrate the controlled formation of localized light emitters in AlN thin films grown on SiC by molecular beam epitaxy (MBE). In contrast to other works,\cite{xue2020single, xue_experimental_2021, cannon_polarization_2023, pezzagna_photoinduced_2026} we do not observe a detectable number of light emitters in the as-grown AlN thin films. The photoluminescence spectra only exhibit a significant number of sharp emission lines after the AlN thin films have been exposed to a post-growth treatment consisting in proton irradiation followed by an annealing process. Moreover, the density and brightness of the light emitters are much larger in AlN thin films grown under N-rich than under Al-rich conditions, a clear indication that not only the post-growth treatment, but also the growth conditions are a relevant factor for the efficient formation of these localized light sources.


The monocrystalline AlN films investigated in our work were grown on semi-insulating 4H-SiC(0001) substrates using a plasma-assisted MBE technique.\cite{Dinh2025Sep, Yuan2026} Before being loaded into the ultrahigh vacuum environment, the substrates were first cleaned with 30\% HCl to remove the surface oxide as well as surface contaminants, rinsed with de-ionized water, blown dry with a nitrogen gun, and finally outgassed for two hours at 500~$^\circ$C in a load-lock chamber attached to the MBE system.

\begin{figure}
    \includegraphics[width=0.8\columnwidth]{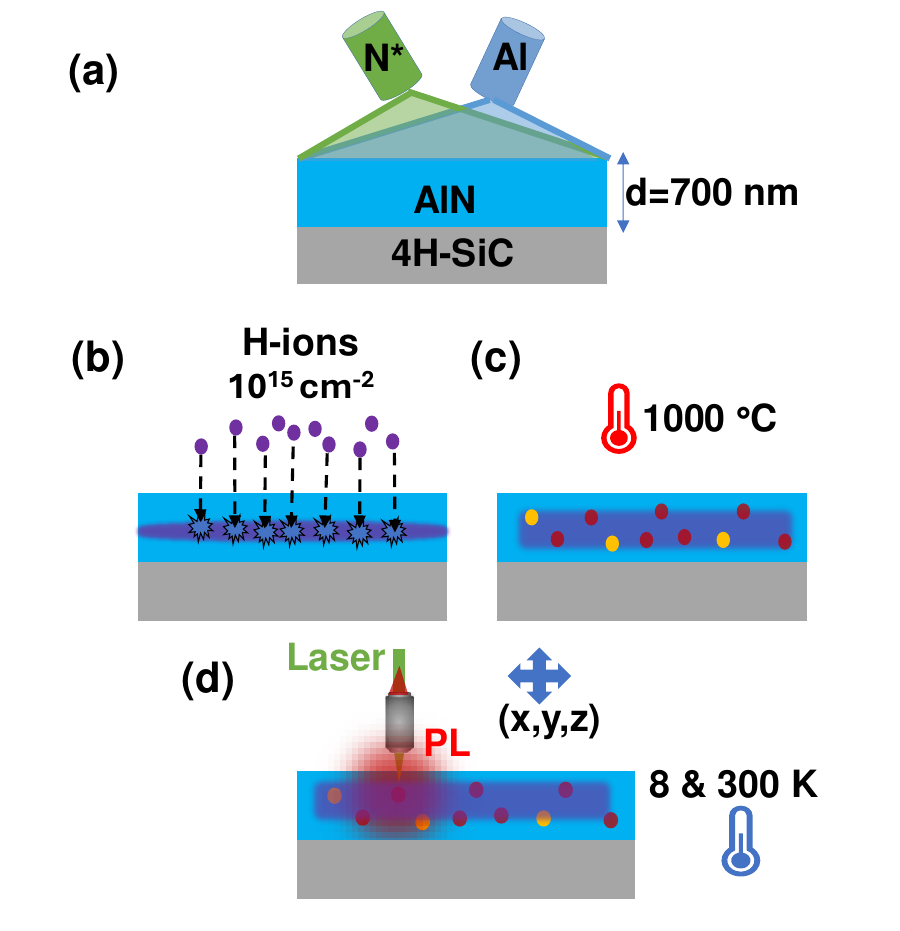}
    \caption{\label{fig:Schematic}
    (a) Plasma-assisted MBE growth of 700-nm-thick AlN on 4H-SiC.
    (b, c) Post-growth formation of light emitters via proton irradiation and/or annealing (left and right panels, respectively).
    (d) Spatially resolved PL characterization of the AlN thin film.
    }
\end{figure}

The MBE-growth chamber is equipped with a high-temperature effusion cell to provide 6N-pure Al metals. A Veeco UNI-Bulb radio-frequency plasma source is used for the supply of active nitrogen (N$^*$). 6N-pure N$_2$ gas is used as N precursor, which is further purified by a getter filter. The N$^*$ flux is calculated from the thickness of a GaN layer grown under Ga-rich conditions, and thus with a growth rate limited by the N$^*$ flux.\cite{Dinh2023Apr} The 700-nm-thick AlN films used in this work were grown at a thermocouple temperature of 1000~$^\circ$C under either N-rich ($\mathrm{Al/N}^* \approx 0.3$) or Al-rich ($\mathrm{Al/N}^* \approx 1.1$) conditions, see Fig.~\ref{fig:Schematic}(a).

After growth of the epitaxial AlN thin films, the $10\times10$~mm$^2$ samples were cut into multiple pieces, and each of them was exposed to a different combination of the proton irradiation and annealing treatments shown Figs.~\ref{fig:Schematic}(b) and \ref{fig:Schematic}(c). The broad-beam proton irradiation (Ion Implanter High Voltage Engineering Europa B.V., Model B8385) was performed at room temperature using a fluence of $10^{15}$~cm$^{-2}$, an energy of 60~keV, and an incidence angle of $7^{\circ}$ to prevent ion channeling. The energy was selected to generate the maximum density of point defects [nitrogen ($\mathrm{V_N}$) and aluminum ($\mathrm{V_{Al}}$) vacancies] at the middle thickness of the AlN thin films, that is, approximately 350~nm below the top surface, see Sec. S2 of the supplementary material. The annealing was performed in a furnace in which the sample was heated up to 1000~$^\circ$C for 2 hours in Ar atmosphere.
 
The optical characterization was performed in a micro-photoluminescence ($\mu$-PL) spectroscopy setup (Renishaw inVia) equipped with a 532-nm-wavelength laser for optical excitation, see Fig.~\ref{fig:Schematic}(d). The samples were optically excited with a laser power of about 20~mW. In the room-temperature experiments, the laser beam was focused on the sample surface to a spot diameter of about 1.5$~\mu$m using a $100\times$ objective with 0.85 numerical aperture (NA). In the measurements at 8~K, the samples were mounted in a cold-finger cryostat and the laser beam was focused to a spot diameter of about 2.5~$\mu$m using a long-working-distance $50\times$ objective (NA=0.55). In both cases, the PL collected by the objective was spectrally separated from the reflected laser beam before sending it to a monochromator equipped with a charge-coupled device camera for spectroscopic analysis.

\begin{figure}
    \includegraphics[width=0.8\columnwidth]{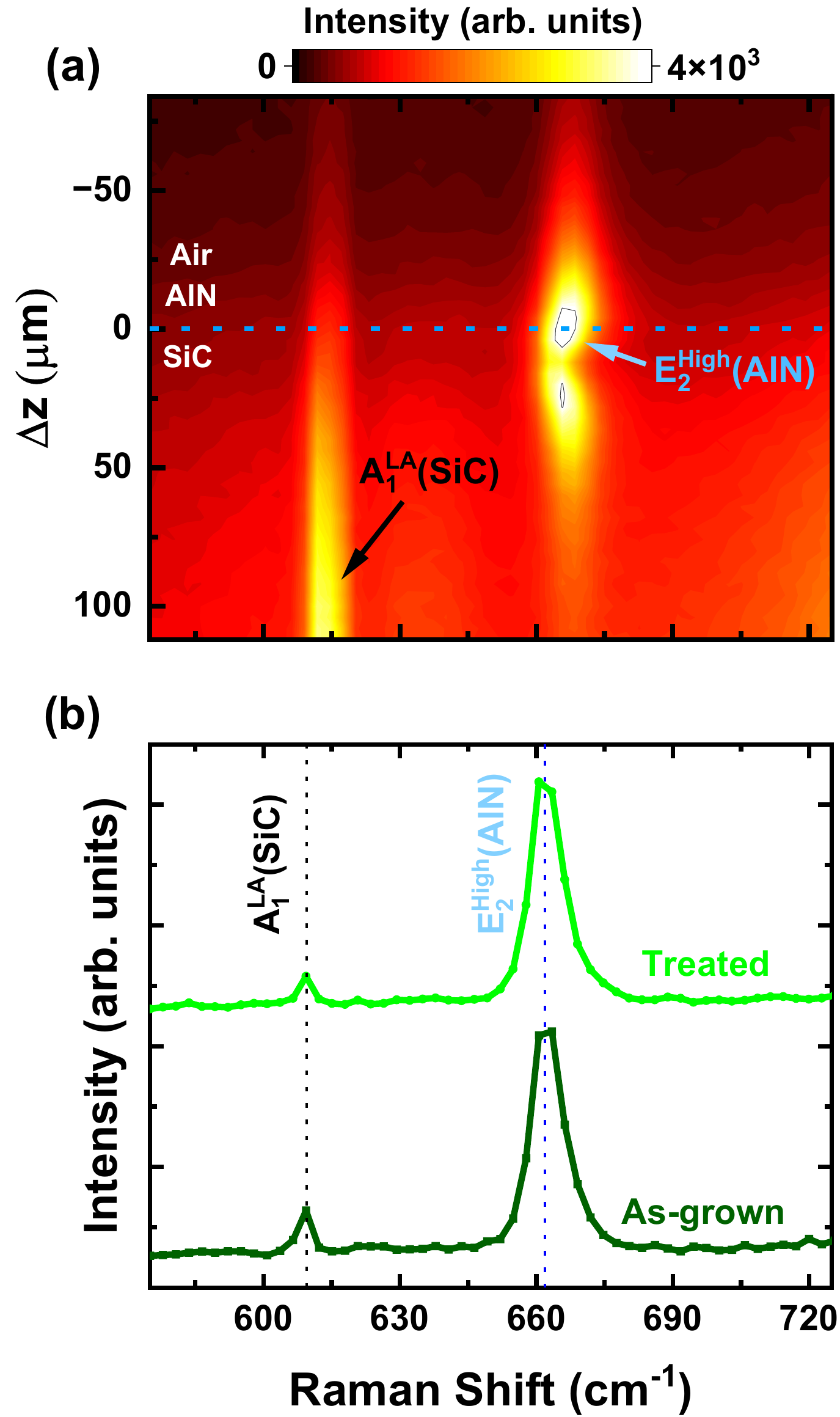}
    \caption{\label{fig:Raman}
    (a) Raman spectrum of the N-rich AlN thin film as a function of the vertical position of the focused laser beam, $\Delta z$. $\Delta z<0$ and $\Delta z>0$ correspond to the focal plane of the laser being above or below the sample surface, respectively.
    (b) Comparison of the Raman spectra measured at $\Delta z=0$ for the as-grown and the post-growth treated AlN (dark and bright green curves, respectively). Both spectra were recorded at room temperature and under the same laser excitation conditions. The curves are vertically shifted for clarity.}
\end{figure}

Before the $\mu$-PL characterization, we first determined the impact of the post-growth treatments on the crystalline quality of the AlN thin films. We did this by comparing the $E_{2}^{high}$ Raman peak of AlN before and after the proton irradiation and annealing treatments. To ensure that the Raman intensities measured at different samples can effectively be compared, we first measured, for each sample, the Raman spectrum as a function of the vertical position of the focused laser beam. Figure~\ref{fig:Raman}(a) shows the result of this vertical scan for the sample grown under N-rich conditions, with the $E_{2}^{high}$ peak of AlN observed at 662$~\wnmbr$ and the $A_{1}^{LA}$ peak of the SiC substrate at $615~\wnmbr$.\cite{bauer_temperature-depending_2009} The intensity of the $E_{2}^{high}$ peak of AlN increases as the focal plane of the laser beam approaches the AlN thin film ($\Delta z<0$), and then decreases as the focal plane moves deeper into the substrate ($\Delta z>0$), where the Raman spectrum is dominated by the $A_{1}^{LA}$ peak of SiC. The Raman spectra used for comparison between samples are those with the maximum intensity of the AlN peak ($\Delta z=0$).

Figure~\ref{fig:Raman}(b) compares the Raman spectra of the AlN thin film grown under N-rich conditions before and after the post-growth treatments (dark and light green curves, respectively). The spectral position, intensity and width of the $E_{2}^{high}$ peaks are similar in both samples, indicating that the proton irradiation and annealing do not have a significant impact on the average crystal quality of the AlN thin film. However, we noticed that the Raman peaks are broader than in AlN thin films reported in other works.\cite{xue2020single} To clarify this behavior, we also performed X-ray diffraction curves of our MBE-grown AlN thin films, see Sec. S1 of the supplementary material. They show that the crystal quality of the MBE-grown thin films depends on the $\mathrm{Al/N}^*$ ratio, with the 0002 reflection peak of the AlN thin film grown under Al-rich conditions being significantly narrower than that of the one grown under N-rich conditions.

\begin{figure*}
    \includegraphics[width=0.8\textwidth]{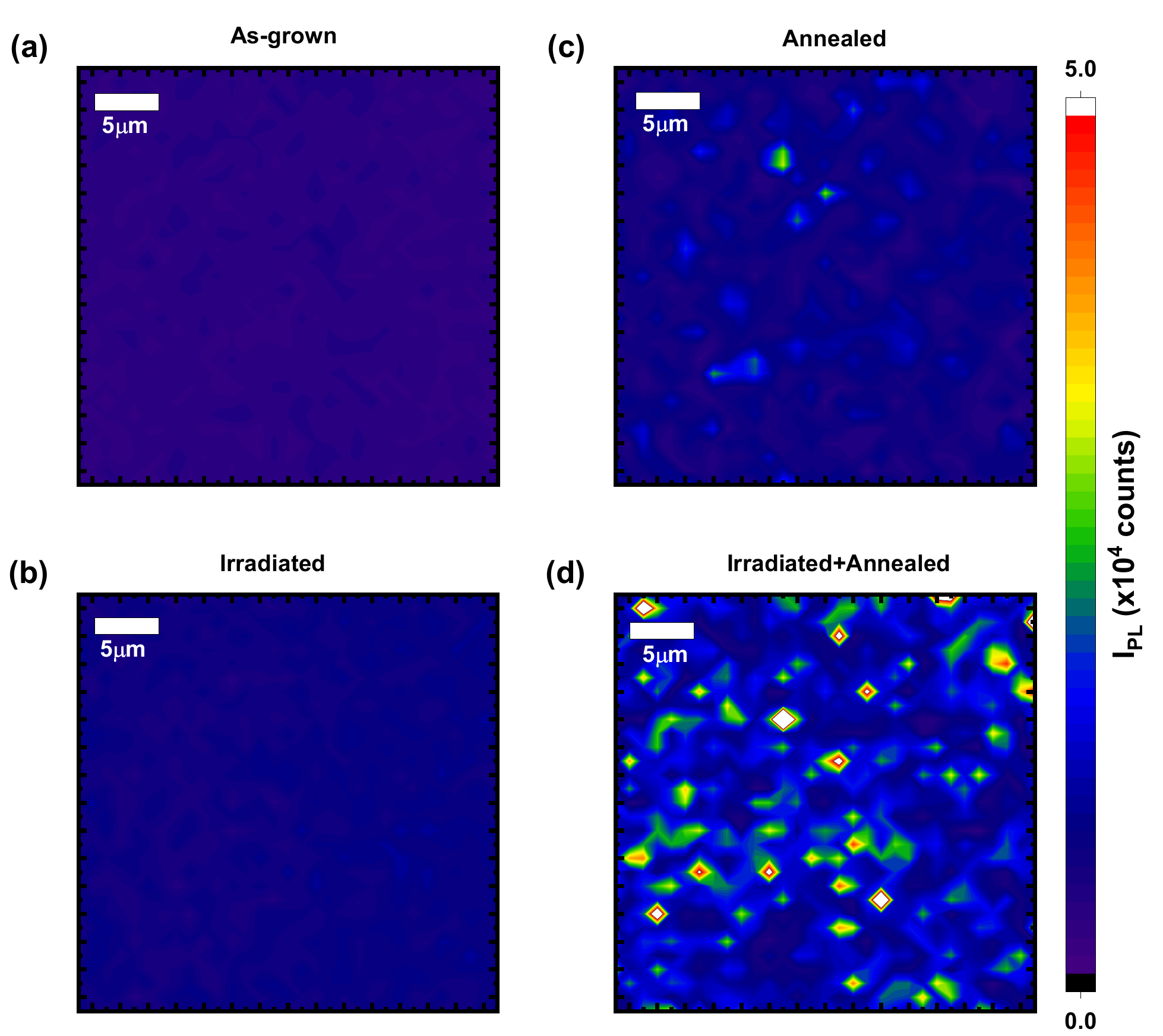}
    \caption{
    \label{fig:2Dmap}
    Panchromatic PL intensity maps of the AlN thin film grown under N-rich conditions, recorded by integrating the PL spectra over the 640--720~nm wavelength range. The maps were measured after the following processing steps: (a) after MBE growth, (b) only proton irradiation, (c) only annealing, and (d) proton irradiation and annealing. All maps were collected at room temperature and under the same laser excitation conditions.}
\end{figure*}

To determine the presence of localized light emitters, we performed panchromatic PL intensity ($I_{PL}$) maps over a $30\times30~\um^2$ area for the as-grown AlN thin films, as well as after each post-growth treatment. They were obtained by integrating the PL spectra recorded at each sample point over the 640--720~nm wavelength range. To ensure that the collected PL is effectively emitted by the AlN, the laser beam was focused at the same vertical position as for the Raman measurements shown in Fig.~\ref{fig:Raman}(b). Figure~\ref{fig:2Dmap} compares the PL intensity maps of the AlN thin film grown under N-rich conditions. In contrast to other works,\cite{xue2020single, xue_experimental_2021, cannon_polarization_2023, pezzagna_photoinduced_2026} the PL intensity of our MBE-grown AlN film on SiC is homogeneous over all the scanned areas, with no evidences of any localized light emission, see Fig.~\ref{fig:2Dmap}(a). A similar result is obtained for the proton-irradiated sample, while the sample that was only annealed shows a few localized points with PL intensities above the background signal, cf. Figs.~\ref{fig:2Dmap}(b) and ~\ref{fig:2Dmap}(c), respectively. A different behavior is observed for the sample piece that was subsequently exposed to both proton irradiation and annealing treatments, see Fig.~\ref{fig:2Dmap}(d). In this case, the spatial map shows many localized points with PL intensities well above the background signal. Assuming that each of these points is associated to a different emission center, we estimate a density of light emitters on the order of $8\times10^{7}$~cm$^{-2}$ for this sample.

\begin{figure}
    \includegraphics[width=0.8\columnwidth]{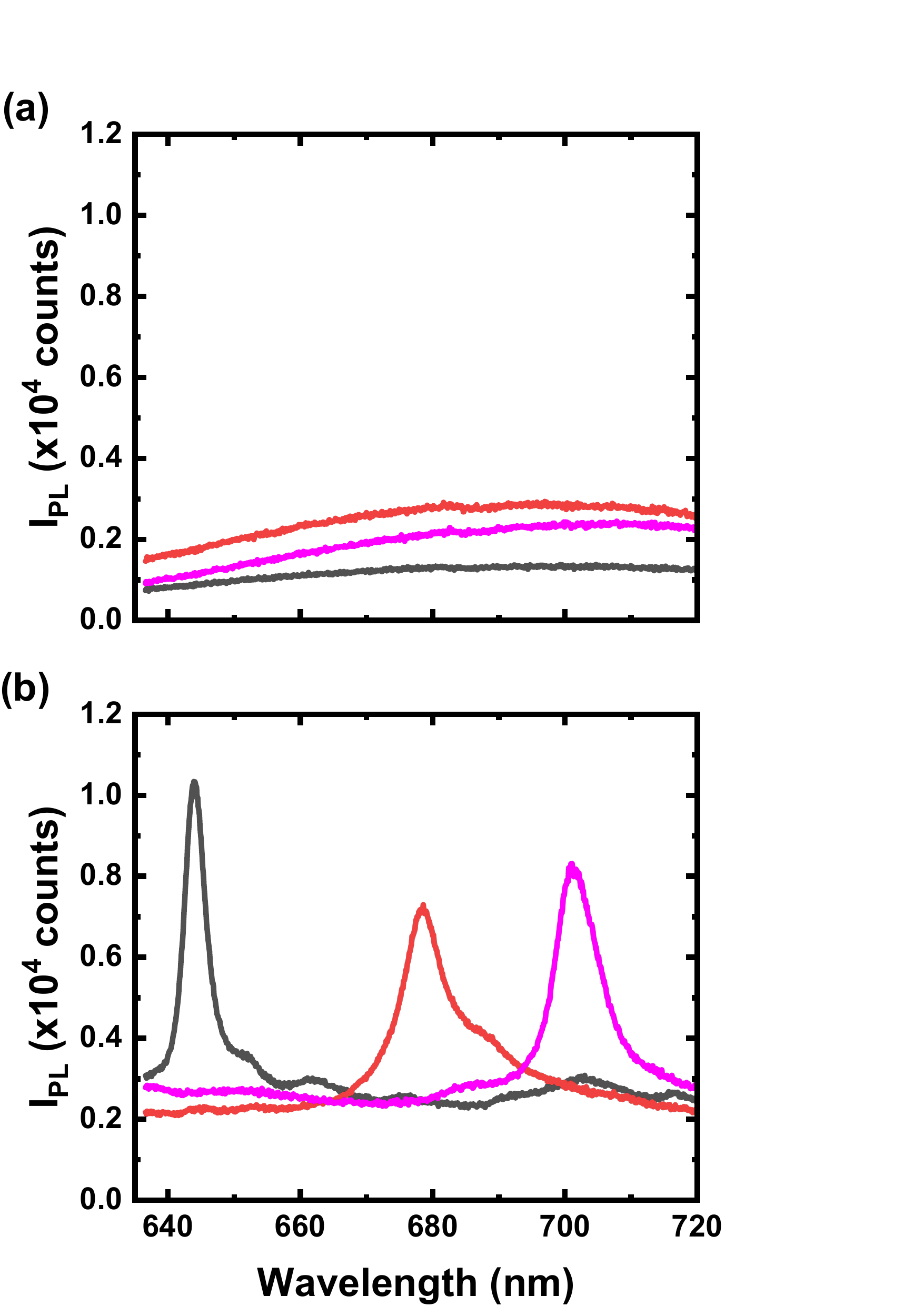}
    \caption{
    \label{fig:Comp}
    (a) Representative PL spectra of the as-grown AlN thin film, measured at three different positions in the spatial map of Fig.~\ref{fig:2Dmap}(a).
    (b) Representative PL spectra of the proton irradiated and annealed AlN thin film, measured at three different positions in the spatial map of Fig.~\ref{fig:2Dmap}(d).
    All measurements were performed at room temperature and under the same laser excitation conditions.}
\end{figure}

Figure~\ref{fig:Comp} compares further the optical behaviors of the as-grown and the fully treated AlN thin films by displaying PL spectra acquired at several positions in each sample. Figure~\ref{fig:Comp}(a) shows three representative PL spectra of the as-grown AlN thin film, each of them consisting of just one broad emission band centered around 700~nm. This broad PL spectrum is observed within the same vertical range of the focused laser beam as in the $E_{2}^{high}$ Raman peak of AlN in Fig.~\ref{fig:Raman}. We therefore attribute it to defect complexes incorporated into the crystal lattice of AlN during the MBE growth process, probably containing vacancy defects and oxygen atoms.\cite{sedhain2012nature, weinstein2012thermoluminescence, zhou2020below, lamprecht2017model}

Figure~\ref{fig:Comp}(b) shows three representative PL spectra of the proton irradiated and annealed sample. Each spectrum contains at least one emission peak within the measured wavelength range, superimposed to the broad emission band already observed in Fig.~\ref{fig:Comp}(a). The wavelengths, intensities and widths of these peaks vary at different positions on the sample surface. This behavior is similar to that of emission centers reported in AlN grown with vapor deposition methods,\cite{xue2020single, xue_experimental_2021, cannon_polarization_2023} thus suggesting that the localized light emitters formed by the post-growth annealing of the MBE-grown films have a similar compositional nature as those observed in AlN thin films grown via vapor deposition methods.

\begin{figure}
    \includegraphics[width=0.8\columnwidth]{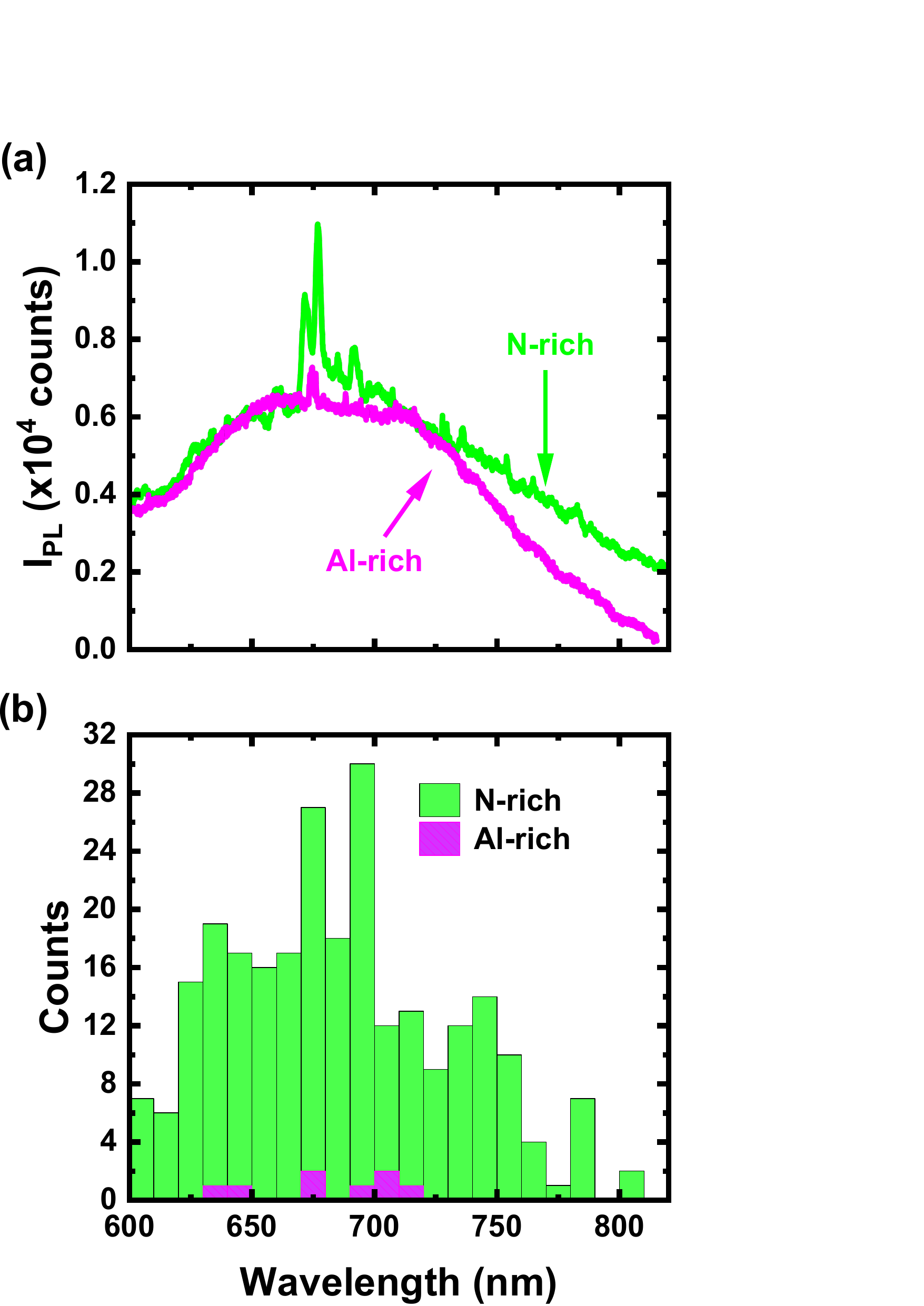}
    \caption{\label{fig:Stats}
    (a) Comparison of representative PL spectra for AlN thin films grown under N-rich (green curve) and Al-rich (violet curve) conditions, after they have been proton irradiated and annealed. 
    (b) Histogram of localized light emitters detected in an area of $50 \times 50~\um^2$, distributed with respect to their emission wavelengths. The green (purple) columns denote the amount of light emitters detected in the sample grown under N-rich (Al-rich) conditions. All measurements were performed at 8~K and under the same laser excitation conditions.
    }
\end{figure}

Finally, we have compared the generation efficiency and optical properties of light emitters formed in the AlN thin films grown under N-rich and Al-rich conditions. Figure~\ref{fig:Stats}(a) displays representative PL spectra of both types of samples, measured at 8~K after proton irradiation and annealing. Both spectra exhibit an emission peak centered at 675~nm, superimposed to the broad emission band already observed in the as-grown AlN films. The line widths of the localized light emitters reduce from approximately 6~nm at room temperature to below 2~nm at 8~K. Interestingly, the localized light emitter in the AlN thin film grown under N-rich conditions is about five times brighter than its counterpart grown under Al-rich conditions. This larger brightness of the light emitters contained in the sample grown under N-rich conditions was consistently observed over the full scanned area, which also exhibited a much larger density of light emitters than in the Al-rich counterpart. This fact is illustrated in Fig.~\ref{fig:Stats}(b), which displays the amount of localized light emitters detected within the 600--800~nm wavelength range in an area of $50 \times 50~\um^2$. We could easily detect more than one hundred light emitters in the AlN thin film grown under N-rich conditions, but less than ten in the Al-rich counterpart (green and violet columns, respectively).

The results reported in Fig.~\ref{fig:Stats} were obtained in AlN thin films that had been both proton irradiated and annealed. Although Figure~\ref{fig:2Dmap} demonstrates that the sequential combination of both treatments is essential for the successful optical activation of the light emitters, the fact that their density and brightness depend significantly on the $\mathrm{Al/N^*}$ ratio used in the MBE chamber indicates that the growth conditions also play an important role in their formation processes. As a matter of fact, theoretical calculations show that the formation energies of intrinsic point defects can depend significantly on the $\mathrm{Al/N^*}$ ratio.\cite{fara_theoretical_1999, gorczyca_theory_1999, stampfl_theoretical_2002, zhu_formation_2024} Moreover, we cannot exclude the contribution of extrinsic impurities, e.g. O, Si and C, which can also be incorporated into AlN during growth.\cite{singhal_molecular_2022} Therefore, a possible scenario compatible with our experimental results could be that, during the annealing process, certain point defects preferentially formed during AlN growth under N-rich conditions combine with $\mathrm{V_N}$ or $\mathrm{V_{Al}}$ centers created during proton irradiation\footnote{according to our calculations, the irradiation with protons generates a similar proportion of $\mathrm{V_N}$ and $\mathrm{V_{Al}}$, see Sec. S2 of the supplementary material} to form the defect complexes responsible for the observed light emitters.

After discussing the formation of localized light emitters in AlN thin films, we briefly discuss the broad PL band spanning from the visible to near-infrared range. Several studies have attributed this broad emission to defect complexes such as $V_{\mathrm{Al}}^{-m}$ and $V_{\mathrm{Al}}-nO_{\mathrm{N}}^{-m}$, with $m$ and $n$ denoting the charge state and number of incorporated O atoms, respectively.\cite{sedhain2012nature, weinstein2012thermoluminescence, koppe2016overview, lamprecht2017model, zhou2020below} We have indeed observed that this broad emission is about ten times brighter in the AlN thin films grown under N-rich than Al-rich conditions (see Sec. S3 of the supplementary material) in agreement with the fact that the growth of AlN under N-rich conditions favors the formation of Al vacancies.\cite{fara_theoretical_1999, gorczyca_theory_1999, stampfl_theoretical_2002, zhu_formation_2024} However, we have also noticed that the subsequent post-growth treatments tend to cancel this intensity difference, probably due to the partial passivation of the Al vacancies by the irradiated protons,\cite{limpijumnong_passivation_2001, maki_identification_2011} leading to a similar intensity of the broad PL in both types of AlN thin films, see Fig.~\ref{fig:Stats}(a). We expect that the further tuning of the $\mathrm{Al/N^*}$ ratio used in the MBE growth, together with the optimization of the parameters used in the post-growth treatments, will minimize and eventually suppress this broad background emission, which is an important condition for the successful application of the localized light emitters as single photon sources.


In conclusion, we have demonstrated the controlled formation of localized light emitters in AlN thin films grown on SiC by molecular beam epitaxy. The PL intensity and spectroscopy maps of the as-grown samples do not reveal localized light emitters in the visible to near-infrared wavelength range. Strikingly, a density on the order of $8\times10^{7}$~cm$^{-2}$ is observed in the AlN thin films grown under N-rich conditions, after the films have been exposed to a post-growth treatment consisting in proton irradiation followed by annealing. In contrast, the same treatment performed in AlN grown under Al-rich conditions leads to a much lower density of light emitters. This different behavior highlights the role of the growth conditions in the subsequent formation of these light emission centers. A possible scenario compatible with our results is the formation of the optically active defect complexes during the annealing as a combination of $\mathrm{V_N}$ or $\mathrm{V_{Al}}$ centers created during the proton irradiation process with point defects preferentially formed during AlN growth under N-rich conditions.

These results provide important insights toward the realization of integrated quantum photonics in AlN-based platforms. As an example, the further tuning of the AlN growth conditions, in combination with focused proton irradiation, could lead to the controlled formation of localized light emitters at well-defined positions within planar optical waveguides or photonic cavities. Moreover, the controlled formation of single photon emitters and spin centers in AlN thin films could also lead to integrated electro-opto-mechanical architectures\cite{liu2023aluminum} for quantum computing and sensing applications similar to those already reported in other solid-state quantum systems.\cite{choquer_quantum_2022, groll_topical_2026, krenner_2026_2026}


\section*{Supplementary Material}
The supplementary material supports the presented results by providing information about X-ray diffraction measurements of as-grown AlN thin films, proton irradiation calculations, and the effect of the post-growth treatment on the broadband PL emission.


\begin{acknowledgments}
The authors would like to thank Carsten Stemmler for technical assistance with the MBE\#9 system, Georg Hoffmann for assistance with the annealing furnace, and Oliver Brandt for a critical reading of the manuscript. D.V.D (Project No. 563156864) and M.Y. (Project No. 563184308) thank the support by the Deutsche Forschungsgemeinschaft (DFG) within the Priority Programme SPP2477 “Nitrides4Future -- Novel Materials and Device Concept". M.S, D.V.D, M.Y. and A.H.-M. acknowledge financial support from the BMFTR and the Senate of Berlin. Support from the Ion Beam Center at Helmholtz-Zentrum Dresden-Rossendorf for the proton irradiation is gratefully acknowledged.
\end{acknowledgments}


\section*{Data availability}
The data that supports the findings of this study are available from the corresponding author upon reasonable request.


\providecommand{\noopsort}[1]{}\providecommand{\singleletter}[1]{#1}%

\end{document}